\documentclass[sigchi]{acmart}

\setcopyright{none} 
\acmSubmissionID{} 
\acmDOI{} 
\acmISBN{}
\acmYear{} 
\copyrightyear{} 
\acmSubmissionID{}
\acmConference[Arxiv]{Online}{2025} 

\usepackage{subfig}

\AtBeginDocument{%
  \providecommand\BibTeX{{%
    \normalfont B\kern-0.5em{\scshape i\kern-0.25em b}\kern-0.8em\TeX}}}

\begin{document}

\title{Vibr-eau: Emulating Fluid Behavior in Vessel Handling through Vibrotactile Actuators}

\author{Frank Wencheng Liu, Ryan Wirjadi, Yanjun Lyu, Shiling Dai}
\author{Byron Lahey, Assegid Kidane, Robert LiKamWa}
\affiliation{
  \institution{Arizona State University}
  \city{Tempe}
  \country{United States}
}

\renewcommand{\shortauthors}{Liu et al.}

\begin{abstract}
Existing methods of haptic feedback for virtual fluids are challenging to scale, lack durability for long-term rough use, and fail to fully capture the expressive haptic qualities of fluids. To overcome these limitations, we present Vibr-eau, a physical system designed to emulate the sensation of virtual fluids in vessels using vibrotactile actuators. Vibr-eau uses spatial and temporal vibrotactile feedback to create realistic haptic sensations within a 3D-printed vessel. When the users are in the virtual environment and interact with the physical vessel, vibration impulses are triggered and the user will feel like there is fluid in the vessel. We explore the impact of motor density, direct touch, and vibration strength on users' perception of virtual fluid sensations. User studies reveal that Vibr-eau effectively simulates dynamic weight shifts and fluid-like sensations, with participants reporting experiences closely resembling real-world interactions with fluids. Our findings contribute to the development of adaptable and scalable haptic applications for virtual fluids, providing insights into optimizing parameters for realistic and perceptually faithful simulated fluid experiences in VR environments.
\end{abstract}



\maketitle

\section{Introduction}

\begin{figure}[t]
\centering
\includegraphics[width=.85\columnwidth]{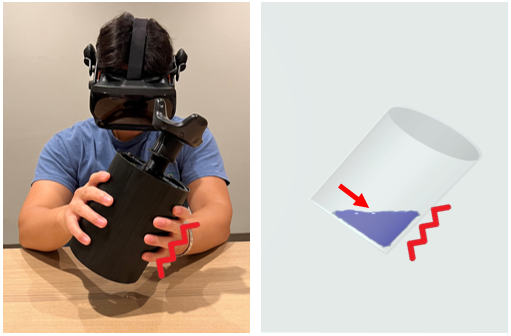}
\caption{Vibr-eau uses multiple vibrotactile actuators to emulate the haptic sensation of virtual liquid in a container. Left: User holding Vibr-eau system and feeling vibrations when virtual liquid hits the side of container. Right: What the user sees in virtual space.}
\label{fig:introductionfig}
\end{figure}

Human-computer interaction (HCI) research in the realm of virtual environments has made significant strides in simulating realistic sensations to enhance user experiences. One of the challenges revolves around the faithful reproduction of fluid sensations in virtual environments. SWISH \cite{swish} and Geppetteau \cite{gepp} offer two different approaches of dynamically shifting weight to render haptic sensation of virtual fluids behavior. However, actively moving systems are more susceptible to failure with frequent use, making them less reliable in demanding environments. In dynamic settings like schools, for example a highschool chemistry class, where equipment is subject to constant handling and rough use, durability becomes critical for ensuring long-term functionality and minimizing maintenance.

Harnessing vibrotactile actuators to emulate the sensation of virtual fluids enhances the viability of haptic systems for real training/classroom use and opens new avenues for exploration. Vibrotactile actuators, known for their accessibility, small form factor, and durability, have been instrumental in simulating various haptic illusions, such as bending, stretching of rigid objects \cite{pseudobend}, stiffness \cite{softness, vibero}, forces \cite{hapcube, impacts}, and textures \cite{textureswithvibrotactile}. Research work have also shown asymmetric vibrations to provide the sensation of direction \cite{waves, slidingvibration, Traxion}, and pseudoforces \cite{grabity, 7463151} typically implemented using voice coil actuators. There is an emphasis on the importance of lightweight form factors \cite{hapticslitreview} and simplicity \cite{torc} for haptic devices, highlighting their significance for user comfort and long-term usage.

Prior research work have explored creating virtual fluid applications; however, they fall short in the haptic rendering of virtual fluids. DualVib \cite{dualvib} was able to use vibrations to simulate the sensation of dynamic mass for solid virtual objects such as coins in a jar. However, the user study results indicated that DualVib did not work out for fluids. This may be a result of the one-axis pseudo-force feedback not being able to capture the full expressiveness of fluid. Another work used multi-actuator vibrotactile feedback \cite{multiactuator} to simulate a wine in a bottle. The authors simulated fluid inside a virtual object using an underdamped spring attached to a virtual mass at the object's center, causing lag and oscillation during movement pauses. The vibration amplitude on each actuator was inversely proportional to the distance from the mass, creating a tactile sensation of motion between two actuator locations. However, this 1D approach fails to capture the complex behaviors of real liquids. 

We present Vibr-eau, a physical system designed to simulate the sensation of virtual fluids in vessels through integration with game engine software. Vibr-eau uses up to 8 vibrotactile motors, arranged in a circular array inside of a 3D-printed vessel. These motors actuate at different times and locations, thus having variations in the tactile feedback that result in asymmetric sensations that mimics the dynamic behavior of liquid. When users interact with the vessel in a virtual environment, vibration impulses are triggered as the center of gravity of the virtual liquid aligns with the container's side and surpasses a predefined acceleration threshold. This approach, grounded in kinesthetic forces and surface vibrations during liquid-container collisions, makes users feel as if there is fluid in the vessel.

With the Vibr-eau system, our goal is to create adaptable, scalable, and general-purpose haptic applications for virtual fluids, addressing a gap in previous research that has not utilized 2D vibrotactile actuators to emulate virtual liquids. We investigated how motor density, direct touch, and vibration strength affect user perceptions of dynamic weight shifts and fluid-like sensations in a virtual container. Our findings identified the optimal parameters for the best user experience within our exploration space. We conducted perceptual studies with 16 participants, comparing Vibr-eau with real liquid and a static haptic proxy, and tested its adaptability to different vessel shapes. Users frequently associated Vibr-eau's tactile sensations with liquid in a container, demonstrating its effectiveness in simulating virtual fluids. Vibr-eau illustrates a method for using vibrotactile patterns to emulate virtual fluids, aiming to inspire future research in this area.
\section{Related Works}

\subsection{Handheld Haptic Devices}
Commerical VR controllers have are both portable and easy to use. However, many of these off-the-shelf controllers only have basic vibrotactile feedback for virtual object contact. Many different haptic hand held interfaces have been developed to enhance the sensations felt in the virtual space. Haptic Revolver \cite{hapticrevolver} allows for the functionality of VR controllers but also renders texture and contact of various objects using a configurable wheel on the device. CLAW \cite{claw} augments traditional controllers with force feedback and actuated movement of the index finger. PaCaPa \cite{pacapa} and CapstanCrunch\cite{capstancrunch} are controllers with movable arms that can produce touch sensation, grasp force feedback, and object textures through vibration. HaptiVec \cite{haptivec} modifies controllers so a user can feel directional haptic pressure vectors. X-Rings \cite{xrings} proposes shape-changing controllers. While these devices provide enhanced sensations in the virtual space, these systems involve complex mechanical structures which make their generalizability limited.

\subsection{Vibrotactile Sensation}
Vibrotactile motors have been used in research to create a wide range of different haptic illusions. Some vibrotactile haptic illusions include bending, stretching of rigid objects \cite{pseudobend}, stiffness \cite{softness, vibero}, compliance\cite{kooboh, 7556272, 3dpress, reflex, hapthimble, curvebutton}, forces \cite{hapcube, impacts, 8816116, 1407031, Traxion, distinctpseudoattraction, 7463150, 7463151, waves}, and textures \cite{textureswithvibrotactile, touchmover}. TORC \cite{torc} presents a rigid haptic controller that renders virtual object characteristics and behaviors such as texture and compliance. Grabity \cite{grabity} has shown that asymmetric vibration and skin stretch can  simulate different levels of percieved weight.
DualVib \cite{dualvib} utilized vibrations to emulate the perception of dynamic mass for virtual solid objects, such as coins in a jar. Despite its effectiveness in solid simulations, user study results revealed its limitations in reproducing realistic sensations for virtual fluids. The authors mention DualVib's limitation of rendering the haptic sensation of virtual fluids as a result the system's one-axis pseudo-force feedback. A separate work employed multi-actuator vibrotactile feedback \cite{multiactuator} to replicate the sensation of wine in a bottle. The vibration actuation followed a spring oscillation pattern back and forth. However, this one-dimensional methodology falls short in accurately capturing the nuanced and expressive behaviors characteristic of real liquids. Both methods of one-dimensional vibrotactile feedback fall short of charactizing the haptic sensations of virtual fluids. There remains an opportunity to explore how we might use multiple vibrotactile actuators to represent the haptic sensation of virtual fluids.

\subsection{Dynamic Weight Shift}

The exploration of dynamic weight shift in the context of haptic feedback has garnered attention in recent research. TorqueBAR \cite{torquebar} was a device that could change its center of mass along 1 degree of freedom. As the user tilted the device, the centre-of-mass shifted in real-time according to a computer-controlled algorithm.
Shifty \cite{shifty} could shift a weight along its main axis to change its rotational inertia and defined the concept of Dynamic Passive Haptic Feedback. 
Researchers have delved into diverse weight-shifting mechanisms, utilizing wind \cite{aeroplane, thorshammer}, air resistance \cite{dragon}, liquid \cite{gravitycup, vibroweight, metaltransfer}, string-driven systems \cite{gepp}, rack and pinion mechanisms \cite{swish}, and vibration \cite{dualvib}. Weight shift applications extend beyond shape simulation \cite{transcalibur, comflex} to include the replication of virtual liquid sensations \cite{swish, gepp, ElastOscillation, dualvib}.

However, a significant number of these existing systems exhibit mechanical complexity, making replication challenging. Some are characterized by bulkiness and discomfort during prolonged use, raising the need for more accessible and user-friendly designs in the exploration of dynamic weight shift in haptic feedback systems. 
Vibr-eau presents a mechanically straightforward solution, utilizing vibrotactile actuators for a user-friendly and mechanically simple haptic feedback system.

\section{Implementation}
\begin{figure}[h]
\centering
\includegraphics[width=1\columnwidth]{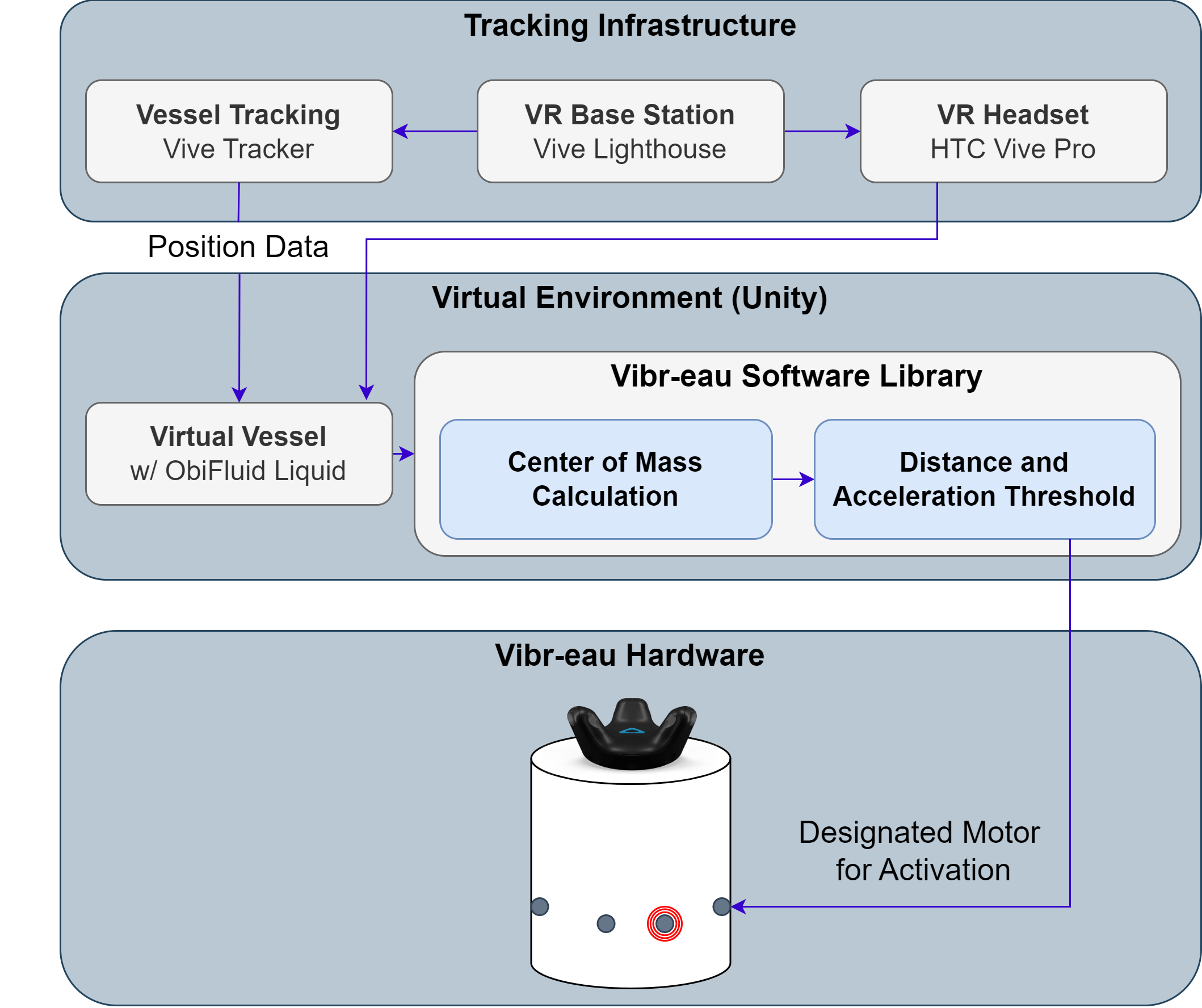}
\caption{Vibr-eau's System Diagram}
\label{fig:systemdiagram}
\end{figure}

In this section, we discuss the components of our implemented Vibr-eau system to promote the reproducibility of our work. Our system diagram is shown in Figure ~\ref{fig:systemdiagram}.

\subsection{Hardware Design}

Our system consists of three components: (i) the Vive tracker (89g) and (ii) the Vibr-eau electronics, which incorporates a NANO 3.0 compatible controller soldered on a perma-proto board (iii) and a physical vessel container. 

\begin{figure}[h]
\centering
\includegraphics[width=.85\columnwidth]{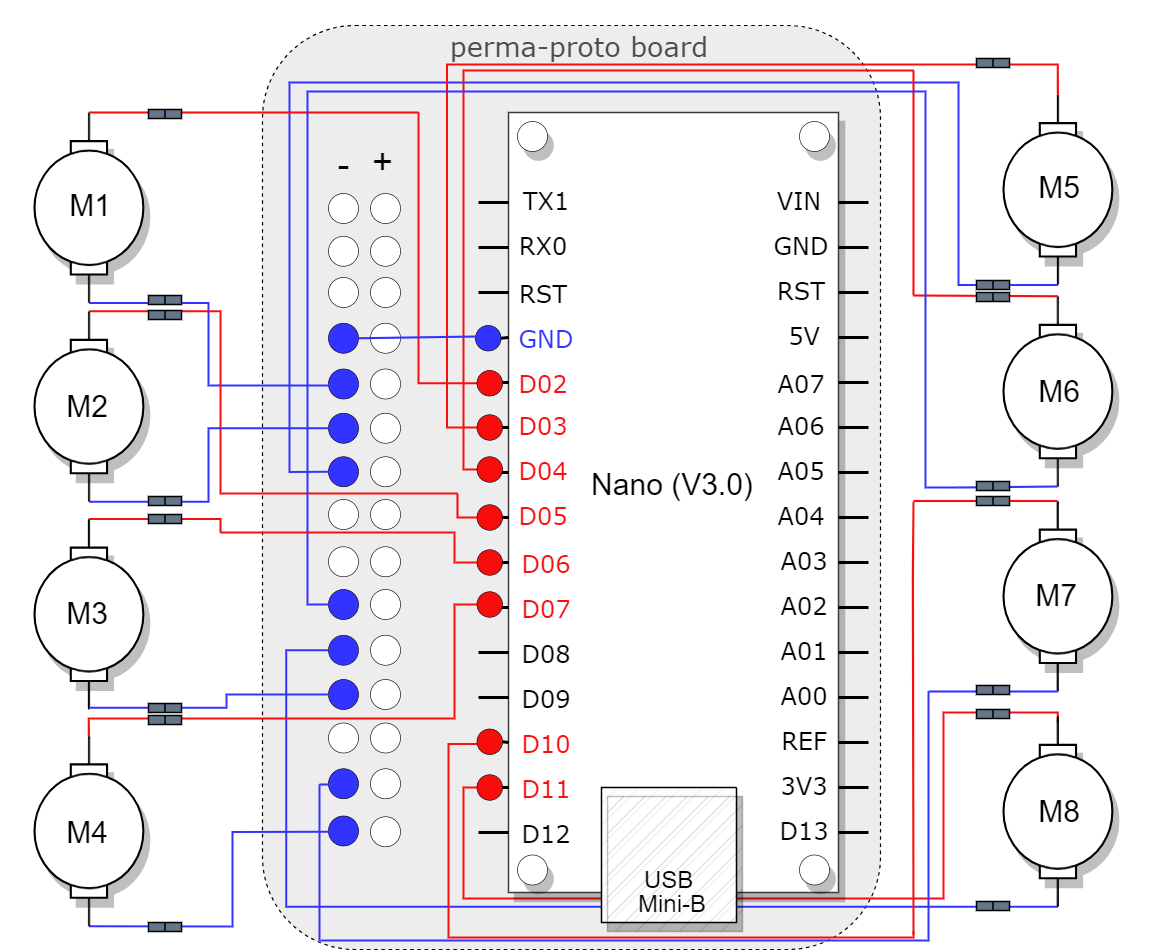}
\caption{Electronic Schematic: a Nano board is mounted on a perma-proto board. Up to 8 vibration motors with male jumper cables can be connected to the 8 female jumper cables extending from the nano board.}
\label{fig:hardwarefig}
\end{figure}

In the following, we describe our electronics (ii). Vibr-eau supports 8 vibration motors and is capable of utilizing a Bluetooth RF transceiver. For our implementation, we opted for a serial port connection via a USB cable. A 3D printed container, created using UltimakerS5, is used to contain the electronics. It has dimensions of 5.8x3.9x2.7cm and a total weight of 6g, making it easily embeddable inside various shapes of fluid containers, such as cylinders, beakers, and spheres. We chose to use flat coreless vibration motors \footnote{Tatoko B07Q1ZV4MJ}, each measuring 10mm x 3mm and weighing at 2g. These motors have a rated frequency of 200hz and rated current of 85mA. According to the spec sheet, these coin motors have a maximum vibration amplitude of 1.2g. The positive wire of each motor is connected to a digital PWM pin, and the negative wire of each motor is linked to the ground pin on the Nano (shown in Figure \ref{fig:hardwarefig}). 
The internal resistance of the motors was measured to be 15.2 ohms. At 5V, there is a total power draw of 1.64 watts per motor. The PWM pins enable precise control over the voltage delivered to the motor, allowing for fine-tuned modulation of vibration strength. The motors are attached on the sides of the physical vessel. The physical vessel can be 3D printed or an arbitrary everyday vessel. 

The aggregate weight of the Vibr-eau electronics encompasses 18g, accounting for the nano, protoboard, and female jumper cables, along with an additional 6g for the 3D printed casing, and an increment of 2g for each added motor with male jumper cables. The maximum total weight of an 8-motor Vibr-eau system weighs 42g. An 8-motor Vibr-eau system inside our cylindrical 3D printed beaker weighs a 270g. The overall weight of components i,ii, and iii weighs a total of 371g.

We had three vessels printed. The cylindrical beaker has a height of 165 mm and a radius of 62.5 mm. The erlen flask also has a height of 165 mm, with a bottom radius of 62.5 mm and a top radius of 20 mm. The Florence flask, with a height of 165 mm, features a bottom radius of 62.5 mm, a middle radius of 80 mm, and a top radius of 20 mm. All containers have a shell thickness of 2 mm.

\subsection{Software Implementation}

For our implementation, Vibr-eau runs using the Unity game engine. 
For our fluid simulations, we chose to use Obi Fluid for real-time fluid dynamics. We chose to use Obi Fluid because it is optimized for Unity and exposes different fluid parameters. We can change the amount of fluid available as well as the fluid properties such as changing the viscosity.

When we interact with water in a water bottle and shake the bottle from side to side, we'll feel the mass of the liquid hit the side of the container. Generally the liquid in containers tends to move together as a cohesive mass. We'll usually feel the mass of liquid hit the side of the container when the shaking is fast. Slow side to side movements, we usually don't feel anything in the container. 

\begin{figure}[t]
\centering
\includegraphics[width=1\columnwidth]{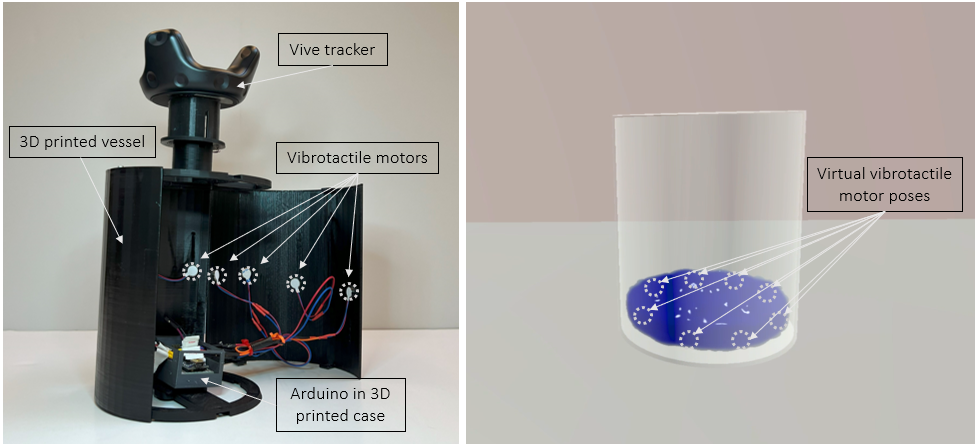}
\caption{Left: The Vibr-eau system comprising of the Vive Tracker, electronics, and a 3D-printed physical vessel, with motor placement highlighted in light gray. Right: The virtual vessel filled with liquid, with invisible game objects strategically positioned on the bottom of the vessel to correspond with their respective physical motors.}
\label{fig:insidefig}
\end{figure}

Drawing inspiration from real-life experiences with handling liquids in containers, we sought to capture the kinesthetic force of liquid movement and the surface vibration between liquid and container collision. Leveraging the observation that liquids typically move cohesively, we posited that the center of gravity (CoG) of the liquid could effectively serve as a proxy for its collective motion. In the virtual environment, we have the virtual fluid spawned into a virtual vessel identical to the physical vessel. The virtual vessel is a .obj file that maps to the same shape of the physical vessel. We then calculate the position of the CoG for virtual fluid particles. We strategically positioned invisible game objects at the bottom of the virtual vessel within the virtual space. We placed vibration motors in a circular array pattern near the middle, offset from where the invisible game objects were, along the side of the physical container. This arrangement (shown in Figure~\ref{fig:insidefig}) aimed to optimize the area where users could perceive sensations, as the virtual liquid remained predominantly near the bottom of the virtual vessel. The Unity game engine would then send commands to the Vibr-eau system so that the vibrotactile motors were activated when the CoG closely approached the virtual game object and surpassed a predetermined acceleration threshold. Users would feel the motors actuated pulses on fast shakes, swirls and movements. The motors would not activate on slow motions. Additionally, we synchronized all the motors to vibrate together when the virtual liquid moved from the bottom to the top and then back down again in the vessel. This was done to simulate the realistic behavior of fluid for both upward and downward shaking movements. As the binary vibrotactile actuators actuate to follow the center of mass of virtual liquid, haptic asymmetry is introduced through spatial and temporal variation. These vibrations would occur on transient events as a result of the virtual fluid, and previous research \cite{verrotouch} has shown that playing vibrations during transient events can render realistic contact sensations. This methodology aims to replicate the behavior of fluid motion, aligning the virtual experience with real-world expectations. 

We incorporated flexibility into our system by enabling developers to program adjustable distance and acceleration thresholds based on their preferences. To determine suitable acceleration parameters for our implementation, we simulated 10 minutes of virtual fluid movement at various speeds and intensities. The acceleration threshold was then selected based on the 25th percentile of acceleration values, ensuring a well-calibrated and responsive virtual experience. The center of gravity (CoG) coordinates were transformed into the local coordinate system of the virtual container. Acceleration thresholds were measured at specific percentiles: 25\% at 0.00001, 50\% at 0.0004, 75\% at 0.00018, and 90\% at 0.0006. Unity uses default units of meters (m). We chose the distance parameter to be 1cm away. All of the properties to control the fluid behavior are adjustable by modifying Obi Fluid parameters in Unity.

\section{Configuration}

To understand the impact of physical liquids on container surfaces and the resultant acoustic signals, we strategically affixed contact microphones to three distinct vessels—a plastic water bottle, a metal cup, and a glass flask. Our approach involved measuring the impact sounds generated by varying water quantities (20g, 50g, and 80g) across different motions, swaying, swirling, and shaking. To measure the physical signal lengths, we conducted each action three times across three actions, three weight values, and three container types, resulting in a total of 81 signals. These measurements were performed manually by the same person to ensure consistency. We aimed to perform consistent "fast" actions. The signal length was determined by measuring the difference between the first and last zero crossings of the signal.

\begin{figure*}[h]
\centering
\includegraphics[width=2\columnwidth]{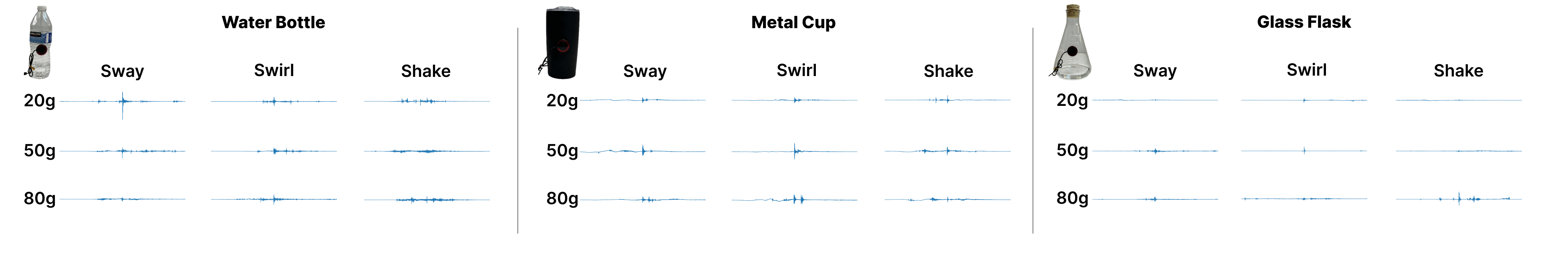}
\caption{Single contact microphone recording: Acoustic vibration samples of single action recorded with a contact microphone in the form of digital audio signals for a plastic water bottle, a metal cup, and a glass flask. All samples are 400 ms long and in the same amplitude (y-axis) scale}
\label{fig:singlemic}
\end{figure*}

\begin{figure*}[h]
\centering
\includegraphics[width=2\columnwidth]{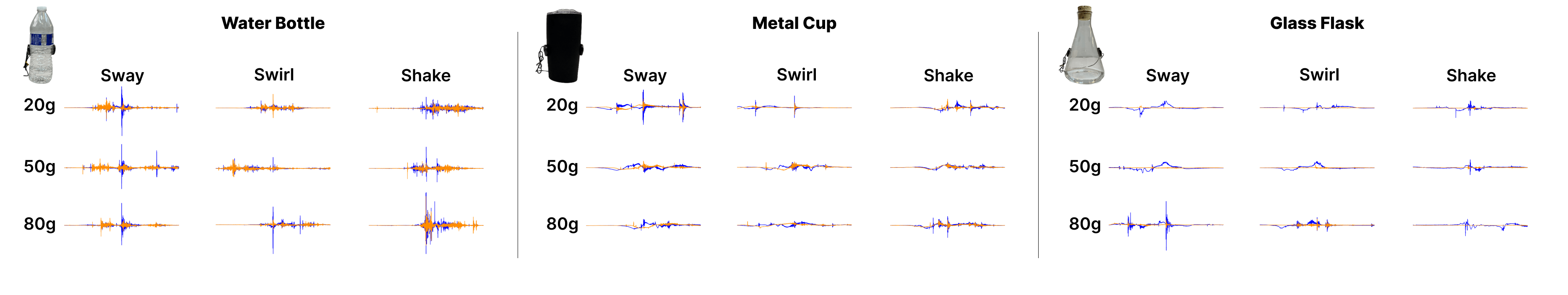}
\caption{Two contact microphone recording: Acoustic vibration samples of single action recorded with two contact microphones in the form of digital audio signals for a plastic water bottle, a metal cup, and a glass flask. All samples are 400 ms long and in the same amplitude (y-axis) scale. Blue for left, Orange for right.}
\label{fig:duomic}
\end{figure*}

In the initial phase, we collected signal data using a single contact microphone, aiming to capture the nuanced experience of an individual holding the vessel and perceiving the resulting impact vibrations. Subsequently, we extended our exploration by employing two microphones to delve into potential spatial aspects, depicted in Figures ~\ref{fig:singlemic} and ~\ref{fig:duomic}. These captured signals assumed a pivotal role in shaping the intricate vibration patterns and duration of the vibrotactile motors integrated into our Vibr-eau system.

Using a Tascam US-2x2 audio/midi interface and Audacity, we visually represented the sound effects corresponding to actions such as sway, shake, and swirl, each associated with different vessels and quantities of water. Each impact signal shown in Figures~\ref{fig:singlemic} and Figure~\ref{fig:duomic} have a duration of 400 milliseconds. Processing the sinlge mic signals, we measured the length of each impact duration. This analysis revealed that the average length of each physical impact signal to be 69.95 milliseconds.

Using the determined average impact duration as our starting point, we conducted a series of experiments, exploring various vibrotactile durations. After conducting personal exploration and considering the feedback from our research team, we determined that changing the duration of the signal to 80ms would likely enhance the user experience. This adjustment was made based on our collective expertise and a desire to optimize the haptic feedback provided by the system. As a result, the duration of the pulses from our Vibr-eau system was set to 80ms.

We employed two contact microphones positioned on opposite sides of the liquid-filled containers. Our analysis revealed that the side of the container struck by the liquid exhibited higher amplitude readings compared to the side unaffected by the liquid during swaying movements. During shaking motions, both sides of the container exhibited similar amplitudes. These signal recordings support our approach to applying vibrations for sway and swirl movements, as well as the decision to activate vibrations uniformly for shaking motions.

\section{User Study}
We conducted two sets of user studies. The first user study was an exploration user study where we explored the effects of changing different parameters of the Vibr-eau system. The second study was a perceptual study where we compared Vibr-eau to baselines of real liquid and a static haptic proxy. We also wanted to understand how well our system generalized to different vessel shapes. We recruited 16 users ages 18-45. 8 users identified as male. 6 users identified as female. 1 user identified as nonbinary and 1 preferred not to answer. 4 users did not have prior experience with virtual reality before. The participants did the two studies on separate days and received a \$25 starbucks giftcard as compensation. All studies were reviewed and approved by the institutional review board.

\subsection{Exploration User Study}

\begin{figure}[h]
\centering
\includegraphics[width=1\columnwidth]{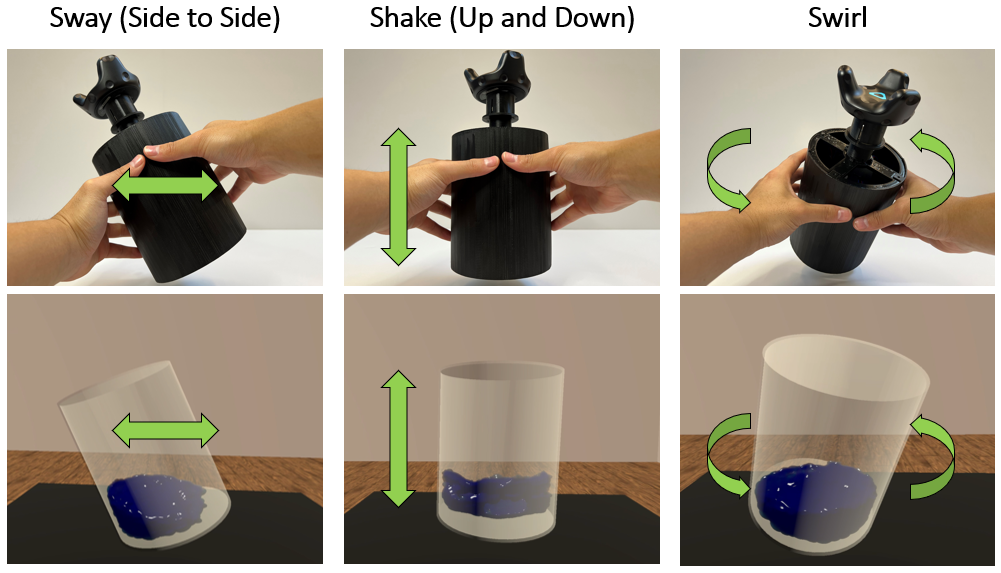}
\caption{Top: The user's point of view while performing sway, shake, and swirl with the physical system. Bottom: The corresponding virtual reality display seen by the user.}
\label{fig:motions}
\end{figure}

In our exploratory user studies, our primary focus was on evaluating the Vibr-eau system in a comprehensive way, with a specific focus on key parameters such as \textit{motor density,} \textit{direct touch}, and \textit{vibration strength}. For each parameter exploration, users were asked to perform three actions three times - sway (side to side), shake (up and down), swirl (around the vertical axis) as shown in Figure~\ref{fig:motions}. The users were then asked to play with the system for a minute. Users were asked to perform both one handed and two handed interactions for each condition. Parameter variation randomized for each user. Participants were asked to provide their feedback on their individual experiences by answering four VRUSE\cite{vruse} questions: Q68 ``I had the right level of control over the system", Q48 ``The system behaved in a manner that I expected", Q8 ``I found the system easy to use" and Q98 ``I would be comfortable using this system for long periods of time". They were also asked open-ended questions\footnote{List of study questions can be found in supplemental material} to describe their experience.

\subsubsection{Motor Density}
\begin{figure}[h]
\centering
\includegraphics[width=1\columnwidth]{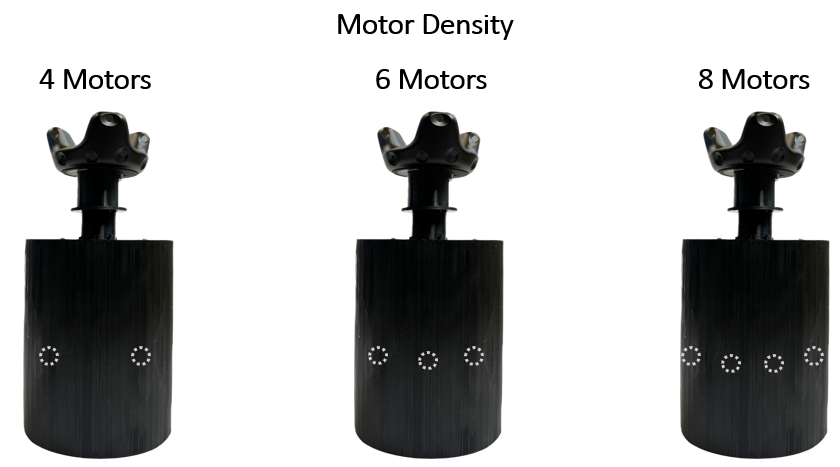}
\caption{From left to right: Frontal view (only half of motors are shown) of the physical systems showcasing varying motor densities: 4 motors, 6 motors, and 8 motors.}
\label{fig:motordensity}
\end{figure}

In this user study, we investigate the influence of motor density on haptic sensation, specifically focusing on the number of motors within the system and the associated spatial patterns. The key question guiding our exploration are centered around understanding how varying the number of motors—specifically, 4, 6, or 8 along the side of a 3D printed container—impacts the overall haptic experience. Figure~\ref{fig:motordensity} depicts the variations.

\textit{Results}. 
Participants often noted discrepancies in feedback intensity and timing, particularly when comparing motions like shaking, swirling, and swaying. While some movements elicited strong and immediate feedback, others lacked responsiveness or were delayed, impacting the overall immersive experience. These discrepancies are reflected in Table~\ref{table:motordensity} with no one condition being better than another. Looking deeper into the user feedback, in the 4-motor condition, 4 users likened the haptic feedback to beans in a container while another mentioned particle dots. In the 6-motor condition, 1 user mentioned beans, while 3 described it as similar to slime or gel. In the 8-motor condition, 1 user compared the sensation to liquid and beans, while 2 felt it resembled liquid in a container. Based on the user responses in the testimonials, it seems like the higher motor density there is the more "liquid" the device feels. 

\begin{table}[h]
\centering
\resizebox{\columnwidth}{!}{%
\begin{tabular}{l|lll|}
\cline{2-4}
                                                                                                                                   & \multicolumn{3}{l|}{Condition}                                                                  \\ \hline
\multicolumn{1}{|l|}{Question}                                                                                                     & \multicolumn{1}{l|}{4 Motors}        & \multicolumn{1}{l|}{6 Motors}         & 8 Motors         \\ \hline
\multicolumn{1}{|l|}{I had the right level of control over the system}                                                             & \multicolumn{1}{l|}{M=4.18, SD=0.88} & \multicolumn{1}{l|}{M=4.13, SD=0.93}  & M=3.875, SD=1.11 \\ \hline
\multicolumn{1}{|l|}{The system behaved in a manner that I expected}                                                               & \multicolumn{1}{l|}{M=3.75, SD=1.03} & \multicolumn{1}{l|}{M=3.5, SD=1.00}   & M=3.5, SD=1.17   \\ \hline
\multicolumn{1}{|l|}{I found the system easy to use}                                                                               & \multicolumn{1}{l|}{M=4.50, SD=1.00} & \multicolumn{1}{l|}{M=4.88, SD= 0.33} & M=4.68, SD=0.58  \\ \hline
\multicolumn{1}{|l|}{\begin{tabular}[c]{@{}l@{}}I would be comfortable using this system \\ for long periods of time\end{tabular}} & \multicolumn{1}{l|}{M=3.88, SD=1.16} & \multicolumn{1}{l|}{M=3.75, SD=1.15}  & M=4.06, SD=0.89  \\ \hline
\end{tabular}
}
\caption{User rating scores for the motor density conditions: 4 motors, 6 motors and 8 motors}
\label{table:motordensity}
\end{table}

\subsubsection{Direct Touch}
\begin{figure}[h]
\centering
\includegraphics[width=.75\columnwidth]{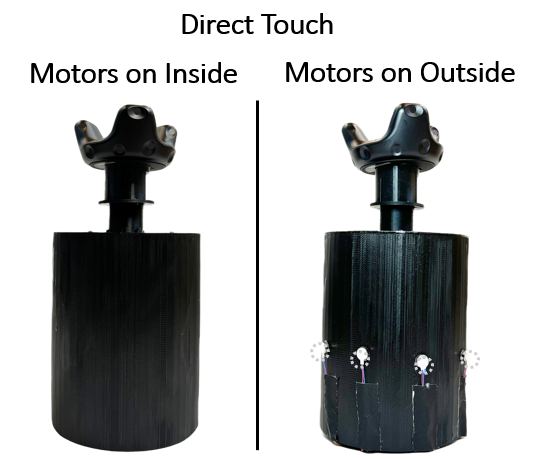}
\caption{Left: Physical system with motors attached on the inside. Right: Physical system with motors attached on the outside.}
\label{fig:directtouch}
\end{figure}

In this user study, we extend our exploration to the realm of direct touch. We aim to discern the impact of tactile feedback when users interact with the motors on the inside attached behind the 3D printed shell and outside the 3D printed shell where users would have direct contact. The key question guiding our investigation is how the user's perception and tactile experience differ when feeling the motors indirectly, behind a material layer, as opposed to direct contact with the motors. Figure~\ref{fig:directtouch} shows the two versions.

\textit{Results}.
In examining the user rating scores from Table~\ref{table:directtouch}, it seems like the motors on the inside performed slightly better on average than motors on the outside for three of the questions. Preferences regarding motor placement varied among participants. Some favored having the motors outside the vessel for a more realistic feel, while others preferred them inside, citing distraction and immersion concerns. With the motors on the outside some users preferred this more as User 5 shares, ``As opposed to what I felt in the previous version, this felt more comfortable due to a more pronounced feel of the motors being outside." While other users did not like them on the outside as User 16 writes, ``The container used with the wires broke the immersion a bit for me because it didn't feel like the object I was holding in VR. It created a disconnect for me. If the 3D object had wires, then maybe it would have been more immersive and less of a disconnect." With the motors on the inside, User 16 contrasts with the motors on the outside writing, ``As I moved the device around, the liquid moved and the device was very responsive. This was more accurate and created a better, and more immersive, experience overall." User 13 contributes, ``I felt that vibrations in this trial matched my movements the best." Overall, there seems to a slight preference to where motors are on the inside of the 3D printed vessel. 

\begin{table}[h]
\centering
\resizebox{\columnwidth}{!}{%
\begin{tabular}{l|ll|}
\cline{2-3}
                                                                                                                                   & \multicolumn{2}{l|}{Condition}                          \\ \hline
\multicolumn{1}{|l|}{Question}                                                                                                     & \multicolumn{1}{l|}{Motors Inside}   & Motors Outside   \\ \hline
\multicolumn{1}{|l|}{I had the right level of control over the system}                                                             & \multicolumn{1}{l|}{M=4.06, SD=0.89} & M=4.00, SD=1.00  \\ \hline
\multicolumn{1}{|l|}{The system behaved in a manner that I expected}                                                               & \multicolumn{1}{l|}{M=3.75, SD=0.83} & M=3.75, SD=0.97  \\ \hline
\multicolumn{1}{|l|}{I found the system easy to use}                                                                               & \multicolumn{1}{l|}{M=4.56, SD=0.60} & M=4.44, SD= 0.70 \\ \hline
\multicolumn{1}{|l|}{\begin{tabular}[c]{@{}l@{}}I would be comfortable using this system \\ for long periods of time\end{tabular}} & \multicolumn{1}{l|}{M=4.06, SD=0.66} & M=3.81, SD=1.01  \\ \hline
\end{tabular}
}
\caption{User rating scores for the direct touch conditions: motors on the inside and motors on the outside of the 3D printed vessel.}
\label{table:directtouch}
\end{table}

\subsubsection{Motor Strength}
\begin{figure}[h]
\centering
\includegraphics[width=1\columnwidth]{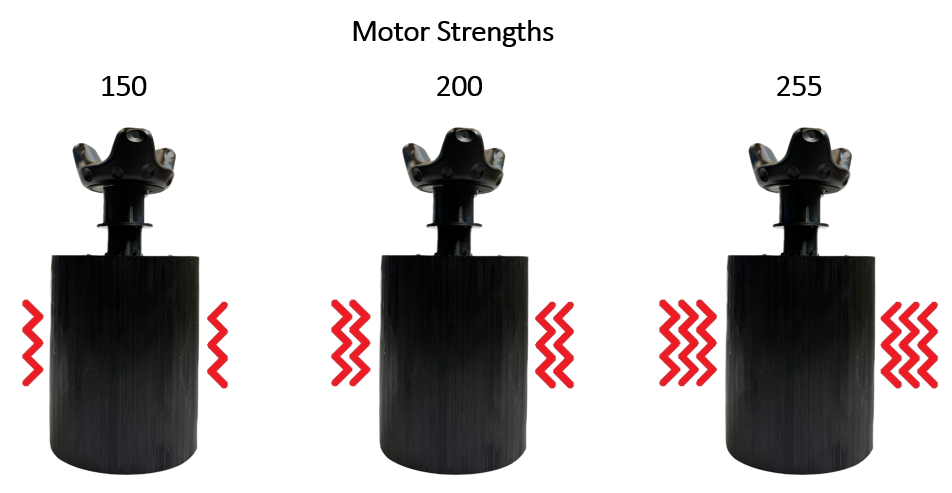}
\caption{From left to right: Physical systems showcasing varying motor strengths: 150, 200, 255.}
\label{fig:vibstr}
\end{figure}

In this user study section, we examine the impact of vibration strength on user experience. Our objective was to understand how varying the strength of the motors influences tactile feedback. Specifically, we investigated whether different levels of vibration strength affect users' perceptions and interactions with the system. Using PWM pins, our range of PWM values spans from 0 (representing 0\% duty cycle, where power is off) to 255 (representing 100\% duty cycle at 5V output). In our study, we tested three PWM values: 150, 200, and 255. Figure~\ref{fig:vibstr} illustrates the different strengths.

\textit{Results}. 
We see from Table~\ref{table:vibstr} that the 255 condition performed higher than the others in 2 questions on average. In examining the user testimonials feedback intensity varied across conditions, with some participants preferring stronger vibrations for better immersion. For example in the 150 condition, User 11 shares, ``Felt like there was moving liquid. The vibration didn't feel too strong, but that made it feel more realistic." While User 16 disagrees, ``Sometimes I could not feel any vibration when there should be one. This broke the immersion a bit. When we had the previous version with the strong vibrations, I was sure to feel everything since it was strong." These sentiments were consistent across conditions 200 and 255 as well. 

Some participants associated the different tactile sensations from the motor strengths with different materials, including liquid, foam, and beans in a container. However, the consistency of these associations varied depending on the specific condition.
In the 150 condition, two users described the sensation as similar to beans, with one user emphasizing a "mechanical movement feeling." Additionally, two users compared the sensation to that of a smartphone.
Moving to the 200 condition, one user mentioned soft foam, while two users likened it to beads, and another user compared it to beans.
In the 255 condition, three users associated the sensation with beans, while one user compared it to the vibrations of a game station controller. 
The remaining participants in these conditions compared the haptic sensation to water or liquid.

\begin{table}[h]
\centering
\resizebox{\columnwidth}{!}{%
\begin{tabular}{l|lll|}
\cline{2-4}
                                                                                                                                   & \multicolumn{3}{l|}{Condition}                                                                 \\ \hline
\multicolumn{1}{|l|}{Question}                                                                                                     & \multicolumn{1}{l|}{PWM 150 (2.94V)} & \multicolumn{1}{l|}{PWM 200 (3.92V)}  & PWM 255 (5.00V) \\ \hline
\multicolumn{1}{|l|}{I had the right level of control over the system}                                                             & \multicolumn{1}{l|}{M=4.13, SD=0.78} & \multicolumn{1}{l|}{M=4.56, SD=0.49}  & M=4.38, SD=0.69 \\ \hline
\multicolumn{1}{|l|}{The system behaved in a manner that I expected}                                                               & \multicolumn{1}{l|}{M=3.88, SD=0.86} & \multicolumn{1}{l|}{M=4.06, SD=0.66}  & M=4.38, SD=0.69 \\ \hline
\multicolumn{1}{|l|}{I found the system easy to use}                                                                               & \multicolumn{1}{l|}{M=4.63, SD=0.48} & \multicolumn{1}{l|}{M=4.63, SD= 0.60} & M=4.63, SD=0.60 \\ \hline
\multicolumn{1}{|l|}{\begin{tabular}[c]{@{}l@{}}I would be comfortable using this system \\ for long periods of time\end{tabular}} & \multicolumn{1}{l|}{M=4.06, SD=0.83} & \multicolumn{1}{l|}{M=4.13, SD=0.78}  & M=4.38, SD=0.78 \\ \hline
\end{tabular}
}
\caption{User rating scores for the motor strength conditions: 150, 200 and 250.}
\label{table:vibstr}
\end{table}

\subsection{User Experience Comparison Study}

In our perceptual user studies, we aimed to evaluate the performance of our Vibr-eau system against real liquid and static haptic proxies, while also assessing its generalization across various vessel profiles. Based on feedback from the exploration user study, we decided to choose 8 motors, with motors on the inside of the 3D printed vessel, at a vibration strength of 255 as the parameters for the perceptual study. For each condition and vessel shape, users were asked to perform three actions three times - sway (side to side), shake (up and down), swirl (around the vertical axis) as shown in Figure~\ref{fig:motions}. Users were then asked to play around with the system for a minute. Users were asked to perform both one handed and two handed interactions for each condition. Conditions and vessel shapes were randomized for each user.  Feedback was collected using a 7-point Likert scale for object realism and open-ended questions \footnote{List of study questions can be found in supplemental material} to capture individual experiences. For analysis of the data, we conducted the Friedman test to determine whether there were statistically significant differences among the groups for object realism scores. If there were sigificant differences (P<0.05), we then used the Conover post-hoc test to find significant differences (P<0.05) between all group pairs. We also share notable user responses.

\subsubsection{Baseline Comparisons}

\begin{figure}[h]
\centering
\includegraphics[width=1\columnwidth]{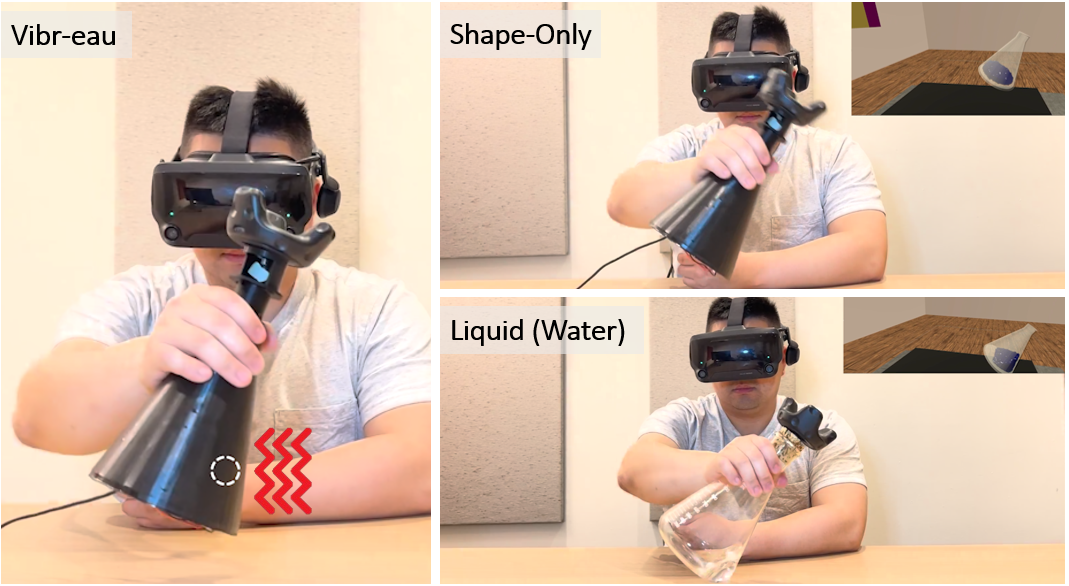}
\caption{User interacting with Vibr-eau system on left. White circles highlight which motors are activated. User interacting with Shape-Only static haptic proxy top right, erlenmeyer flask filled with 80g of water bottom right.}
\label{fig:baseline}
\end{figure}

In this study, our goal was to evaluate the haptic sensations provided by Vibr-eau in an Erlen vessel shape compared to two baselines: a static haptic proxy represented by an identical-shaped vessel without vibrations, and real liquid contained in a 1000ml Erlenmeyer flask filled with 80g of water (405g + 95g Vive tracker). These are shown in Figure~\ref{fig:baseline}. We hypothesized that the haptic feedback generated by our Vibr-eau system would achieve similar object realism to those of real liquid. 

\begin{figure}[h]
\centering
\includegraphics[width=1\columnwidth]{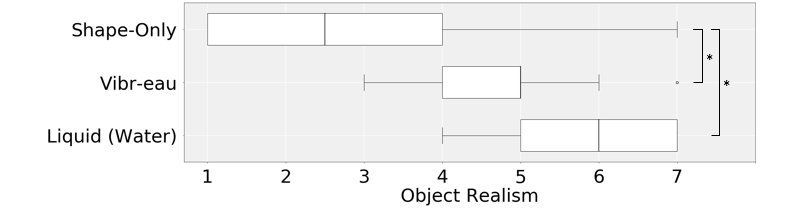}
\caption{Object Realism scores for Shape-Only static haptic proxy, Vibr-eau, and Liquid. Significance marked by asterisk.}
\label{fig:baselineresults}
\end{figure}

\textit{Results}. We found that users rated both the Vibr-eau system (M=4.9, SD=1.0)  and real liquid (M=5.9, SD=1.1) conditions highly in object realism. Users found the shape-only static haptic proxy (M=2.75, SD=1.7) unconvincing. With the Friedman test, we found statistical differences among the three groups. 
There was no statistically significant difference between the mean of the Vibr-eau system and liquid (P=0.092). 
Statistically significant differences were found between Vibr-eau and the Shape-only version (P=0.044) as well as Liquid and the Shape-only version (P<.01). The distribution of data can be seen in Figure~\ref{fig:baselineresults}.

We found that nine out of sixteen users perceived a weight shift when using the Vibr-eau system, indicating a successful emulation of the dynamic mass of virtual fluids through vibrations. User 14 shares, ``I was overall able to feel a shift in weight when slowly and aggressively tilting the object." User 3 writes, ``I was able to feel the weight change, more when I was swaying the beaker's lower section than when I shook the top." While Vibr-eau was able to emulate the weight shift of virtual liquids for some users, on the other hand six users encountered inconsistencies in the system's response, reporting instances where expected motions failed to trigger vibrations or where vibrations were expected but not felt during interactions. 

In the liquid condition, the majority of participants (14 out of 16) reported perceiving a weight shift when interacting with the real liquid. Six users expressed dissatisfaction, noting discrepancies between the behavior of the real liquid and their expectations based on the virtual simulation. User 9 elaborates, ``Honestly the only thing that seems off is that the liquid does not seem to share the same consistency/viscosity of the virtual liquid." 

In the shape-only condition, only two out of sixteen users perceived a weight shift, possibly due to visual illusions. Most participants reported not feeling any haptic feedback.

Overall users found interacting with the Vibr-eau system compelling. User 11 testifies, ``When comparing it to an actual flask with liquid, it's a lot more realistic than I expected." User 4 writes, ``The gradual weight shift and apt haptic sensation produced by Vibr-eau were on par [with liquid] and gave a real feedback of holding a vessel with fluid in it." User 10 shares, ``I liked the representation of the Vibr-eau, because it felt like I was feeling the "liquid" in VR, which I think it could be an innovative thing to implement in the world of virtual reality."

\subsubsection{Multiple Vessels}
\begin{figure}[h]
\centering
\includegraphics[width=1\columnwidth]{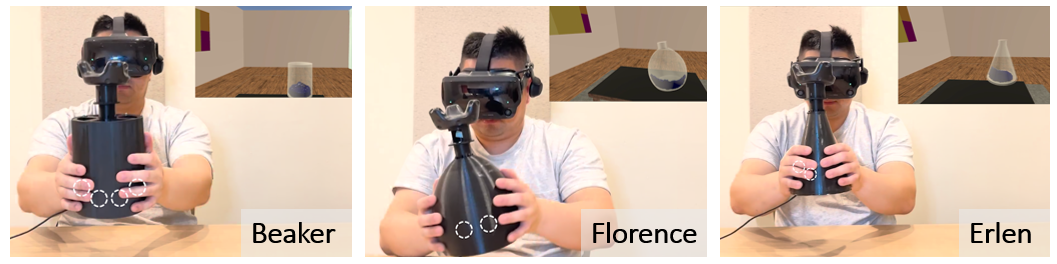}
\caption{From left to right: User interacting with beaker, florence, erlen vessel shapes. White circles highlight which motors are activated.}
\label{fig:multiple}
\end{figure}

In this user study, our exploration expands to encompass diverse shape profiles—namely, the beaker (cylinder), florence (sphere), and erlen (cone) vessel shapes. These are shown in Figure~\ref{fig:multiple}. We hypothesized that the haptic feedback provided by our Vibr-eau system would provide the sensation of virtual fluids consistently across different vessel shapes.

\begin{figure}[h]
\centering
\includegraphics[width=1\columnwidth]{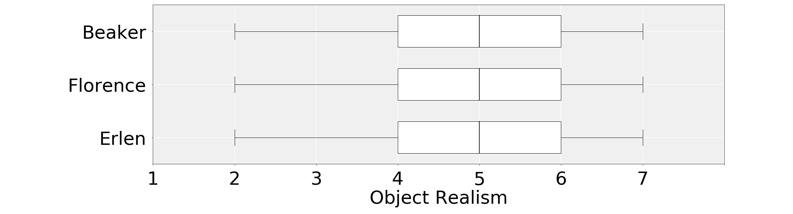}
\caption{Object Realism scores for vessel shapes: Beaker, Florence and Erlen. Significance marked by asterisk.}
\label{fig:multipleresults}
\end{figure}

\textit{Results}. The means for the Beaker(M=4.9, SD=1.4), Florence(M=4.9, SD=1.4), and Erlen(M=4.8, SD=1.2) vessel shapes were all fairly consistent. 
In computing the Friedman Test, there was no statistical significance among the pairwise comparisons.
Overall based on these scores, the Vibr-eau system across different vessel shapes performed similarly in regards to object realism scores. The distribution of data can be seen in Figure~\ref{fig:multipleresults}.

Users reported mixed experiences with haptic feedback, with some finding it accurate and others noting inconsistencies, particularly during gentle movements. Users reported feeling weight shift for different vessel shapes: 7/16 for the beaker, 7/16 for the Florence, and 7/16 for the erlen. We believe that different hand sizes may have impacted the user experience. User 13 shares, "Overall, I think I had a preference for the other shapes compared to the sphere. The sphere I thought was too big and loose to handle comfortably. The cylinder I found better to handle than the sphere and had a better shifting of weight to it. Overall, I thought the flask to be the best." Meanwhile User 2 believes that, "I think that the sphere seemed the best. Maybe because of the way I held it? It seemed to have the strongest likeness to the liquid in VR." User 8 writes, "I liked the erlen the most, it was the easiest to handle and was the most responsive. The sphere was the least responsive and the most unwieldy, and the cylinder stands somewhere in the middle." Interactions involving one hand versus two hands, as well as the size of the vessels, all contribute to the user experience, reflecting the variability observed in user responses. Overall, we did find that the Vibr-eau system was able to generate the haptic sensation for virtual fluids across different vessel shapes.
\section{Discussion}
\subsection{Dynamic Weight Shift Perception via Asymmetric Vibration - Spatial and Temporal}
Even though our methodology does not have an actively shifting mass, user testimonials showed that weight shift was perceived as a result of the vibrations. A majority of users were surprised how well vibrations could emulate the sensations of liquid. 13/16 users expressed such sentiments in the testimonials despite the absence of physical liquid. Interestingly enough, it seemed the physical weight of the device also played a role in capturing the realism of virtual liquid. One user mentioned in the user experience comparison study that since the Vibr-eau system weighed less than the beaker of water and as a result its lack of weight took away from the realism. “Vibr-eau can try increasing the weight of the vessel to make it feel more realistic. The vessel in Vibr-eau was so light that I expected to feel harder impact of the liquid on its walls (the vibrations are not the same as how a liquid would impact the walls) whereas the baseline beaker was heavy enough for me to lower my expectations about the impact force of the liquid.”

The results of our study demonstrate the effectiveness of the Vibr-eau system in conveying dynamic weight shifts of virtual liquids through spatially and temporally asymmetric vibrations. We believe the spatial and temporal asymmetry of the vibrations played a crucial role in enhancing the realism of the haptic experience. By activating motors selectively based on the position and movements of the virtual liquid, the system created nuanced tactile sensations that aligned with users' expectations of interacting with a liquid and emulated the sensation of dynamic weight shift.

\subsection{Haptic Rendering of Fluid Behavior}
The qualitative feedback from participants further supports the effectiveness of the Vibr-eau system in simulating liquid-like sensations. Many users explicitly mentioned that the haptic feedback they experienced closely resembled the tactile qualities of liquid in a container. Users also mentioned that Vibr-eau was able to render haptic sensations to other materials in the container such as beans, beads, or foam. This indicates that the Vibr-eau system was successful in eliciting the intended perceptual responses, as users were able to accurately recognize and interpret the haptic sensations as representative of a liquid medium.

\subsection{Improvements and Future Applications}
We also elicited user feedback facilitating future improvement of Vibr-eau as well as asking them to brainstorm potential applications of our system. Participants noted inconsistencies in the intensity and timing of haptic feedback, particularly during slow motions where the motors remained inactive. Future iterations of the system could address these issues by implementing more responsive motor activation algorithms and fine-tuning vibration parameters to ensure a more consistent and immersive haptic experience.

The Vibr-eau system opens up exciting opportunities for various applications across domains such as virtual reality gaming, education, and training. By accurately replicating the tactile properties of different materials within virtual containers, the system can enhance the realism and effectiveness of virtual simulations and training scenarios. Users suggested applying the Vibr-eau system in VR theme park settings, gaming, physical science labs, and remote learning.

\section{Limitations}
\subsection{Inconsistency of Haptic Rendering}
As some users pointed out, there were inconsistencies in the intensity and timing of the haptic feedback across different conditions and interactions.
The vibration rendering and inconsistency of feedback was a limitation of this work. The decision to activate motors based on predetermined acceleration thresholds and distance events, such as fast shakes or swirls, may have resulted in inconsistent timing of haptic feedback. Users who came in with the expectation of always feeling haptic feedback may have experienced delays or inaccuracies in haptic rendering, particularly during slow motions where the motors remain inactive. To mitigate this issue in future iterations, implementing a continuous scaled vibration strength approach could prove beneficial. This method would adjust the strength of vibration based on the speed of the motion, with slower motions triggering lower intensity vibrations and faster motions resulting in stronger vibrations. Additionally, discrepancies between the visual movement of the virtual fluid and the corresponding haptic feedback from the physical system were observed. Aligning the parameters of the virtual fluid to prevent excessively rapid movement could help address this discrepancy and enhance overall coherence between visual and tactile feedback.

\subsection{Dependency on Visual Cues}
Users' perception of the haptic feedback may be influenced by the accompanying visual stimuli or virtual environment. This dependency on visual cues could affect the system's effectiveness. For future work we can explore how the visual aspect of the virtual fluid might impact the perceived haptic feedback.

\section{Future Work}

\subsection{Motor Placement}
In our work, we strategically positioned the motors in a circular array within the 3D printed vessel. An avenue for future work involves examining the influence of various motor placement patterns on user perception. Could specific placements be deemed optimal for vessels of varying shapes? What discernible reactions might users exhibit towards these distinct arrangements?

\subsection{Material Type and Liquid Behavior}
In future work, we plan to manipulate parameters such as frequency, amplitude, and waveform of the motor signals to explore the replication of tactile sensations associated with various materials. Our research will involve experimenting with different vibration strengths across multiple motors. For instance, we aim to adjust the frequency and intensity of vibrations to mimic the viscosity or density of a liquid, while also varying the amplitude and duration of pulses to recreate the texture of solids such as grains or beads. This exploration will provide valuable insights into how different materials can be simulated through haptic feedback, potentially enhancing the realism and versatility of our system.

\subsection{Multi-sensory Integration}
Within the baseline study, 6 users commented how the audio feedback of the real liquid splashing against the container walls added to the realism and immersion. As future work, we plan on integrating visual and auditory cues to enhance the fidelity of material replication. By synchronizing haptic feedback with visual representations of the virtual container and its contents, we can create a more immersive multisensory experience. Additionally, incorporating sound effects that correspond to the tactile sensations experienced could further enhance the illusion of interacting with virtual fluids.

\subsection{Exploring the Impact of Weight}

In the perceptual study, 2 users highlighted how the lightness of the system took away from the realism of Vibr-eau. User 5 writes, ``Vibr-eau can try increasing the weight of the vessel to make it feel more realistic. The vessel in Vibr-eau was so light that I expected to feel harder impact of the liquid on its walls (the vibrations are not the same as how a liquid would impact the walls) whereas the baseline beaker was heavy enough for me to lower my expectations about the impact force of the liquid." User 6 also agrees, writing, ``When compared with the real liquid, Vibr-eau was less realistic. The real liquid container was heavier, making you feel like you were holding glass. Vibr-eau was lighter, and sometimes you could tell you were holding plastic." As future work, we can explore how the weight of the actual device can add or detract from the haptic illusion of the vibrotactile motors.
\section{Conclusion}
Vibr-eau is a system that employs multiple vibrotactile actuators to replicate the haptic sensations of virtual fluids within a container, utilizing spatial and temporal asymmetry. From our studies, users could perceive dynamic weight shifts of virtual liquid through spatially and temporally asymmetric vibrations. Vibr-eau was able to emulate the haptic sensation of liquid nearly as well as physical liquid in a container. Vibr-eau was able to generalize well for different vessel shapes. The overall feedback from users indicates a positive response to the system's ability to deliver liquid-like haptic sensations. Vibr-eau holds promise for applications ranging from gaming to virtual training scenarios, offering users a more immersive and engaging experience when interacting with virtual fluids.

\bibliographystyle{plain}
\bibliography{vibreau}
\nocite{*} 
\end{document}